# *Performance v. Turnover: A Story by 4,000 Alphas*

Zura Kakushadze[§†1] and Igor Tulchinsky[¶2]

[§] *Quantigic® Solutions LLC,[3] 1127 High Ridge Road, #135, Stamford, CT 06905*

[†] *Free University of Tbilisi, Business School & School of Physics
240, David Agmashenebeli Alley, Tbilisi, 0159, Georgia*

[¶] *WorldQuant LLC, 1700 East Putnam Ave, Third Floor, Old Greenwich, CT 06870*

September 7, 2015

Abstract

We analyze empirical data for 4,000 real-life trading portfolios (U.S. equities) with holding periods of about 0.7-19 trading days. We find a simple scaling $C \sim 1/T$, where $C$ is cents-per-share, and $T$ is the portfolio turnover. Thus, the portfolio return $R$ has no statistically significant dependence on the turnover $T$. We also find a scaling $R \sim V^X$, where $V$ is the portfolio volatility, and the power $X$ is around 0.8-0.85 for holding periods up to 10 days or so. To our knowledge, this is the only publicly available empirical study on such a large number of real-life trading portfolios/alphas.

---

[1] Zura Kakushadze, Ph.D., is the President and a Co-Founder of Quantigic® Solutions LLC and a Full Professor in the Business School and the School of Physics at Free University of Tbilisi. Email: zura@quantigic.com

[2] Igor Tulchinsky, M.S., MBA, is the Founder and CEO of WorldQuant LLC. Email: igort.only@worldquant.com

## 1. Introduction

The stock market provides a wide array of opportunities for investing. Investment strategies vary by a number of important characteristics, such as long-only vs. dollar-neutral and in between, average holding period (or, equivalently, turnover), portfolio return, volatility, Sharpe ratio, etc. Holding periods (or horizons) can range from multi-year (e.g., mutual funds and pension funds) to intraday and even sub-(milli)second (high frequency trading, or HFT).

It is natural to ask, how do portfolio return[4] and Sharpe ratio depend on turnover? In long-only strategies with long holding periods[5] we can take, say, monthly returns. For mutual funds, Carhart [1997] finds[6] that turnover negatively impacts the funds' annual abnormal returns and interprets this as owing to the net cost of trading, i.e., turnover does not aid performance.

In this paper, among other things, we address the question of turnover dependence in the context of dollar-neutral quantitative trading strategies with shorter horizons, with holding periods of roughly from intraday to about 20 trading days. By some estimates (see Salmon and Stoke [2010]), about 70% of total trading volume is driven and executed by computer code. Algorithmic trading encompasses most medium frequency trading and all of HFT worldwide.

We analyze empirical data for 4,000 real-life trading portfolios, or alphas,[7] for U.S. equities. To our knowledge, our analysis is the only publicly available empirical study on such a large number of real-life alphas. We attribute this, in part, to the proprietary nature of such data.

In the realm of quantitative trading strategies, a useful performance characteristic is cents-per-share (CPS), which is profit (in cents as opposed to dollars) per share traded. Shorter-horizon quantitative traders generally are well-aware of the simple "rule-of-thumb": the higher the turnover $T$, the lower the CPS $C$. However, what is the precise relationship, $C \sim 1/\sqrt{T}$, $C \sim 1/T^2$, … ? Based on our empirical analysis, we find a simple scaling:

---

[4] In long-only strategies the return is adjusted for the risk-free rate. In dollar-neutral strategies the stock loan rebate is factored in.

[5] E.g., mutual fund strategies based on analyzing the companies' underlying fundamentals.

[6] Carhart [1997] takes mutual fund returns in excess of the 1-month T-bill return, estimates performance relative to his 4-factor model (which is the 3-factor model of Fama and French [1993] plus the 1-year momentum factor of Jegadeesh and Titman [1993]), and studies the relation of this relative performance to 4 explanatory variables, to wit, expense ratio, turnover, total net assets and maximum load fees, via the Fama and McBeth [1973] regression.

[7] Here "alpha" – following the common trader lingo – generally means any reasonable "expected return" that one may wish to trade on and is not necessarily the same as the "academic" alpha. In practice, often the detailed information about how alphas are constructed may not be available, e.g., the only data available could be the position data, so "alpha" then is a set of instructions to achieve certain stock holdings by some times $t_1, t_2, …$

$$C \sim \frac{1}{T} \tag{1}$$

Consequently, we argue below that the portfolio return $R$ has no statistically significant dependence on the turnover $T$. We also find a scaling:

$$R \sim V^X \tag{2}$$

where $R$ is the portfolio return, while $V$ is the portfolio volatility, and the power $X$ is around 0.8-0.85 for holding periods up to about 10 trading days. Apparently, for such holding periods the return predominantly stems from volatility. For holding periods above 10 trading days the power $X$ drops, indicating that some additional factor(s) (e.g., momentum) might be at play.

The remainder of this paper is organized as follow. We discuss our dataset in Section 2. Section 3 gives details of our empirical analysis. We briefly conclude in Section 4.

## 2. Data

In this section we describe our dataset. The data is proprietary to WorldQuant LLC and is used here with its express permission. We provide as many details as we possibly can within the constraints imposed by the proprietary nature of this dataset.

We start from 4,002 randomly selected alphas from a substantially larger sample (whose size is proprietary). All alphas are for trading U.S. equities with overnight holdings. The trading universes for individual alphas are proprietary, but in most part overlap with a typical universe of most liquid U.S. stocks used in algorithmic (statistical arbitrage) trading.

For these alphas we take the annualized daily Sharpe ratio $S$, daily turnover $T$, and cents-per-share $C$. Let us label alphas by the index $i$ ($i = 1, \ldots, N$), where $N$ is the number of alphas. For each alpha, $S_i$, $T_i$ and $C_i$ are defined via

$$S_i = \sqrt{252}\, \frac{P_i}{V_i} \tag{3}$$

$$T_i = \frac{D_i}{I_i} \tag{4}$$

$$C_i = 100\, \frac{P_i}{Q_i} \tag{5}$$

Here: $P_i$ is the average daily P&L (in dollars); $V_i$ is the daily portfolio volatility; $Q_i$ is the average daily shares traded (buys plus sells) by the $i$-th alpha; $D_i$ is the average daily dollar volume traded; and $I_i$ is the total dollar investment in said alpha (the actual long plus short positions, without leverage). More precisely, the principal of $I_i$ is constant; however, $I_i$ fluctuates due to the daily P&L. So, both $D_i$ and $I_i$ are adjusted accordingly (such that $I_i$ is constant) in Equation (4). The period of time over which this data is collected is proprietary.

Next, we weed out the alphas with negative Sharpe ratios: these are "bad" alphas that have not lived up to their expectations. This leaves us with 3,636 alphas. We then remove outliers in each of the three sets $S$, $T$ and $C$ by keeping only the data points that satisfy the condition $|Y_i - \text{median}(Y)| < 3 \text{ MAD}(Y)$, where $Y$ is the placeholder for $\ln(S)$, $\ln(T)$ or $\ln(C)$,[8] and MAD() stands for "median absolute deviation". This gives us $N = 3{,}289$ alphas. The summary[9] of the data (after removing the outliers) is given in Table 1.

All alphas are real-life trading alphas and the performance figures are out-of-sample by definition. There is no survivorship bias in our dataset. The Sharpe ratio and CPS are computed before trading costs, based on the closing prices on each trading day.

## 3. Modeling Performance v. Turnover

Our approach here is to model the dependence of a performance characteristic $Z$, which is a placeholder for the Sharpe ratio $S$, cents-per-share $C$, portfolio return $R$, etc., via a linear cross-sectional regression with the intercept:

$$\ln(Z_i) = a + \sum_{A=1}^{K} b_A \ln(U_{iA}) + \varepsilon_i$$

Here: $a$ is the regression coefficient for the intercept; $b_A$ are the regression coefficients for the $K$ explanatory variables $U_A$ (each of which is an $N$-vector), which can be, e.g., turnover $T$, volatility $V$, etc. (and their number $K$ can be one or more); and $\varepsilon_i$ are the regression residuals.[10]

### 3.1. Cents-per-Share v. Turnover

In Table 2 (also see Figure 2) we summarize the regression of $\ln(C_i)$ over $\ln(T_i)$ (with the intercept), hence the scaling law in Equation (1). Alternatively, if we run the regression of $\ln(C_i T_i)$ over $\ln(T_i)$, the resulting R-squared is very low ($1.7 \times 10^{-3}$), and so is the coefficient for $\ln(T_i)$ (to wit, 0.043 with t-statistic 2.359, while the estimate and t-statistic for the intercept are the same as in Table 2). So, empirically, we have Equation (1): CPS is inversely proportional to the turnover. Based on Table 2 we can write the following approximate empirical formula:

$$\ln(C) \approx -2 - \ln(T)$$

---

[8] We remove the outliers for the logs as $S$, $T$ and $C$ have skewed distributions (especially $T$ and $C$; see Figure 1).

[9] Using the function `summary()` in the R Package for Statistical Computing, http://www.r-project.org.

[10] We use the R functions `summary(lm())` to obtain regression coefficients, t-statistic, etc.

### 3.2. Cents-per-Share v. Sharpe Ratio

In Table 3 (also see Figure 3) we summarize the regression of $\ln(C_i)$ over $\ln(S_i)$ (with the intercept). So, empirically, we have the scaling

$$C \sim S \qquad (6)$$

Based on Table 3 we can write the following approximate empirical formula:

$$\ln(C) \approx -1 + \ln(S)$$

Equations (1) and (6) suggest $C \sim S / T$: regressing $\ln(S_i)$ over $\ln(T_i)$ we get low R-squared ($6.7 \times 10^{-3}$) and a low coefficient for $\ln(T_i)$ ($-0.074$), so the dependence of $S$ on $T$ is weak.

### 3.3. Cents-per-Share v. Turnover and Sharpe Ratio

Thus, in Table 4 we summarize the regression of $\ln(C_i)$ over 2 explanatory variables $\ln(T_i)$ and $\ln(S_i)$ (with the intercept). The results are consistent with those in Tables 2 and 3. The slight drop in the absolute values of the regression coefficients for $\ln(T_i)$ and $\ln(S_i)$ prompts us to explore another explanatory variable constructed via a combination of $S$, $C$ and $T$.

### 3.4. Return v. Volatility

Let us define

$$R'_i = C_i \, T_i = 100 \, \Pi_i \, R_i \qquad (7)$$

Here $\Pi_i = D_i / Q_i$ (see Equations (4) and (5)). The daily portfolio return (see Equation (3))

$$R_i = \frac{P_i}{I_i} = \frac{1}{\sqrt{252}} \, S_i \, \sigma_i$$

The daily return volatility (i.e., the portfolio volatility per dollar invested)

$$\sigma_i = \frac{V_i}{I_i}$$

The ratio $\Pi_i$ has the meaning of the average (over all stocks) stock price for the universe traded by the alpha labeled by $i$. As mentioned above, there is an overlap between the universes traded by different alphas. It is not unreasonable to assume that $\Pi_i$ is approximately uniform across all (or most) alphas.[11] That is, we assume $\Pi_i \approx \Pi$, where the actual value of $\Pi$ is not going to be important in the following. So, we can express the daily portfolio return $R_i$ and the daily return volatility $\sigma_i$ via $R'_i = C_i \, T_i$ and $\sigma'_i = R'_i / S_i$ (see Figure 4 for $R'_i$ and $\sigma'_i$ densities).

---

[11] For our purposes here it would suffice to assume that $\Pi_i$ is not highly correlated with the portfolio volatility $V_i$.

In Table 5 (also see Figure 5) we summarize the regression of $\ln(R'_i)$ over $\ln(\sigma'_i)$ (with the intercept). (The numeric factors differentiating $R'_i$ from $R_i$ and $\sigma'_i$ from $\sigma_i$ only affect the regression intercept.) Thus, we have the scaling given by Equation (2), where the power $X$ is about 0.8.[12] Since the t-statistic for the intercept and R-squared are not very high, it is instructive to run the same regression for the $N$ alphas broken into deciles according to the daily turnover $T_i$. In Table 6 (also see Figure 6 for comparative plots) we summarize the results. In Figure 7 we plot the values of the estimates of the regression coefficient for $\ln(\sigma'_i)$ v. the deciles. We give the summary of these estimates, with and without including decile 1, in Table 7. While the R-squared in Table 6 might not be stellar – apparently, due to the congregation of low $\ln(R'_i)$ values around the value $\ln(\sigma'_i) \approx -2.5$ (see Figure 5), it appears that, except for decile 1, we can interpret the power $X$ in Equation (2) to be around 0.8-0.85. Decile 1 corresponds to the holding periods of above 9-10 trading days. For such holding periods $X$ drops. This may signal that another factor might be relevant for longer holding periods, e.g., momentum. To discern if this is the case would require more detailed proprietary data including the alpha time series.

### 3.5. A Caveat

One caveat in our analysis of the return v. volatility dependence is that it is indirect in the sense that we do not use the actual data for the daily return volatility $\sigma_i$ (which is the same as $\sigma'_i$ above up to a numeric factor) as such data is proprietary; instead, we use the Sharpe ratio $S$ and the product $R'_i = C_i T_i$ in lieu of the daily return $R_i$ (up to a numeric factor) to deduce the volatility $\sigma'_i$. Direct analysis using data for the daily return volatility $\sigma_i$ would be required to ascertain whether our indirect analysis introduces any undesirable biases. Also, since the daily return $R_i$ is related to the product $R'_i = C_i T_i$ we actually use via Equation (7), we assume that $\Pi_i \approx \Pi$ (see above). To see how much error this introduces into our analysis requires more detailed proprietary data for the average stock price for the universe traded by each alpha. We emphasize that this subsection only concerns Equation (2), while Equation (1) is unaffected.

## 4. Conclusions

To our knowledge, our empirical analysis here is the only one of its kind for such a large number of real-life quantitative trading portfolios. How come? First, the field – naturally – is secretive. Second, such data is proprietary. And third, not that many shops have developed or have access to such a large number of alphas. However, we believe that the future of quantitative trading lies in combining tens of thousands, millions and perhaps even billions of ever-fainter and increasingly more ephemeral and subtler alpha signals, which can only be mined by machines analyzing hundreds and even thousands of exabytes of market data (see Tulchinsky [2015]). The number of mutual funds in the U.S. in 2014 was about 9,260 (see, Statista [2015]). In contrast, the number of a priori viable alphas for, say, U.S. equities is many orders of magnitude larger. This is owing to myriad permutations of individual stock holdings in a (dollar-neutral) portfolio of, say, 2,000 most liquid U.S. stocks that can result in a positive return on relatively short time horizons. Furthermore, the universe of these alphas constantly changes due to their ephemeral nature. As a result, quickly-adaptive, sophisticated and technologically well-equipped quantitative trading operations are required to harvest these alphas and combine them via nontrivial algorithms into a unified "mega-alpha", which is then traded with an added bonus of substantially saving on transaction costs due to automatic internal crossing of trades between the not-too-correlated diverse alphas.

---

[12] If we regress $\ln(R'_i)$ over $\ln(\sigma'_i)$ and $\ln(T)$ (with the intercept), the result is almost unchanged as the turnover dependence of $R'_i = C_i T_i$ is weak (see above). Also, in Section 1, for notational and presentational simplicity we used the portfolio return $V$ in Equation (2). Ignoring impact of trading – and recall that $S$ and $C$ are computed before trading costs – the return $R$ does not depend on the investment level, so we actually have $R \sim \sigma^X$.

Our empirical analysis provides the first step in understanding "phenomenological" properties of these alphas. E.g., we find a simple scaling given by Equation (1), whereby CPS (cents-per-share) scales inversely with the turnover, and consequently the portfolio return does not have statistically significant dependence on the turnover. Why is this useful? First, in some cases it may help a portfolio manager decide what kind of alphas to trade (based on the turnover). Second, once the intercept $a$ in our empirical formula $\ln(C) \approx a - \ln(T)$ is determined on a large number of diverse alphas, new alphas can be categorized based on their CPS as compared to the model fit (e.g., by standard deviations from the fit line given the turnover). Similarly, our empirical scaling given by Equation (2) suggests that for holding periods of up to 10 trading days or so, the return predominantly stems from volatility. In some cases this can be useful in choosing how to combine such alphas into the "mega-alpha".

We hope our empirical analysis gives a glimpse into the complex world of the modern and ever-evolving quantitative trading. Combining and trading large numbers of alphas is not only the future of quantitative trading, but it is in fact the present reality.

## Tables

| Quantity | Minimum | 1st Quartile | Median | Mean | 3rd Quartile | Maximum |
|---|---|---|---|---|---|---|
| $S$ | 0.225 | 0.938 | 1.487 | 1.531 | 2.070 | 4.117 |
| $T$ | 0.054 | 0.193 | 0.297 | 0.341 | 0.405 | 1.492 |
| $C$ | 0.021 | 0.247 | 0.493 | 0.684 | 0.882 | 10.86 |

**Table 1.** Summary (using the R function `summary()`) for the annualized daily Shape ratio $S$, daily turnover $T$, and cents-per-share $C$ for the 3,289 alphas discussed in Section 2.

|  | Estimate | Standard error | t-statistic | Overall |
|---|---|---|---|---|
| Intercept | -1.999 | 0.026 | -77.00 |  |
| $\ln(T)$ | -0.957 | 0.018 | -52.95 |  |
| R-squared |  |  |  | 0.460 |
| F-statistic |  |  |  | 2804 |

**Table 2.** Summary for the cross-sectional regression of $\ln(C)$ over $\ln(T)$ with the intercept; see Subsection 3.1 for details. (In the regressions in this and other tables the multiple R-squared and adjusted R-squared are almost the same; we give rounded multiple R-squared.)

|  | Estimate | Standard error | t-statistic | Overall |
|---|---|---|---|---|
| Intercept | -1.050 | 0.014 | -77.73 |  |
| $\ln(S)$ | 1.000 | 0.021 | 48.39 |  |
| R-squared |  |  |  | 0.416 |
| F-statistic |  |  |  | 2342 |

**Table 3.** Summary for the cross-sectional regression of $\ln(C)$ over $\ln(S)$ with the intercept; see Subsection 3.2 for details.

|  | Estimate | Standard error | t-statistic | Overall |
|---|---|---|---|---|
| Intercept | -2.166 | 0.016 | -139.4 |  |
| $\ln(T)$ | -0.889 | 0.011 | -82.64 |  |
| $\ln(S)$ | 0.920 | 0.012 | 77.86 |  |
| R-squared |  |  |  | 0.810 |
| F-statistic |  |  |  | 7018 |

**Table 4.** Summary for the cross-sectional regression of $\ln(C)$ over $\ln(T)$ and $\ln(S)$ with the intercept; see Subsection 3.3 for details.

|  | Estimate | Standard error | t-statistic | Overall |
|---|---|---|---|---|
| Intercept | -0.159 | 0.059 | -2.693 |  |
| $\ln(\sigma')$ | 0.813 | 0.025 | 32.53 |  |
| R-squared |  |  |  | 0.244 |
| F-statistic |  |  |  | 3287 |

**Table 5.** Summary for the cross-sectional regression of $\ln(R')$ over $\ln(\sigma')$ with the intercept; see Subsection 3.4 for details.

| decile | tvr | hold | est.int | est.vol | std.int | std.vol | t.int | t.vol | r.sq | f.stat |
|---|---|---|---|---|---|---|---|---|---|---|
| 0% | 0.054 | 18.52 | --- | --- | --- | --- | --- | --- | --- | --- |
| 10% | 0.109 | 9.137 | -1.201 | 0.399 | 0.218 | 0.087 | -5.514 | 4.560 | 0.060 | 20.80 |
| 20% | 0.166 | 6.026 | 0.030 | 0.811 | 0.173 | 0.071 | 0.173 | 11.39 | 0.284 | 129.8 |
| 30% | 0.217 | 4.602 | -0.011 | 0.809 | 0.198 | 0.083 | -0.057 | 9.757 | 0.225 | 95.19 |
| 40% | 0.259 | 3.858 | 0.081 | 0.906 | 0.191 | 0.080 | 0.423 | 11.27 | 0.280 | 127.1 |
| 50% | 0.297 | 3.367 | -0.550 | 0.667 | 0.192 | 0.081 | -2.873 | 8.248 | 0.172 | 68.03 |
| 60% | 0.338 | 2.957 | -0.138 | 0.869 | 0.245 | 0.102 | -0.563 | 8.509 | 0.182 | 72.40 |
| 70% | 0.380 | 2.631 | -0.032 | 0.910 | 0.203 | 0.086 | -0.157 | 10.60 | 0.256 | 112.5 |
| 80% | 0.440 | 2.274 | 0.132 | 0.993 | 0.183 | 0.079 | 0.722 | 12.50 | 0.323 | 156.2 |
| 90% | 0.607 | 1.646 | -0.199 | 0.767 | 0.174 | 0.076 | -1.141 | 10.11 | 0.238 | 102.3 |
| 100% | 1.492 | 0.670 | 0.066 | 0.933 | 0.132 | 0.059 | 0.499 | 15.91 | 0.436 | 253.1 |

**Table 6.** Summary for the cross-sectional regression of $\ln(R')$ over $\ln(\sigma')$ with the intercept for each decile for the daily turnover $T$; see Subsection 3.4 for details. The columns are as follows: "tvr" gives the $T$ deciles (the first value 0.054 is $\min(T)$); "hold" gives the corresponding holding periods $1/T$; "est.int" is the estimate for the regression coefficient for the intercept; "est.vol" is the estimate for the regression coefficient for $\ln(\sigma')$; "std.int" and "std.vol" are the corresponding standard errors, while "t.int" and "t.vol" are the t-statistic; "r.sq" stands for R-squared; and "f.stat" gives the f-statistic.

| Quantity | Mean | St.dev | Median | MAD |
|---|---|---|---|---|
| w/ Decile 1 | 0.807 | 0.171 | 0.840 | 0.106 |
| w/o Decile 1 | 0.852 | 0.099 | 0.869 | 0.089 |

**Table 7.** Mean, standard deviation, median and MAD for the 10 (including decile 1) and 9 (excluding decile 1) values of est.vol (coefficient for $\ln(\sigma')$) in Table 6. (Also see Figure 7.)

**Figures**

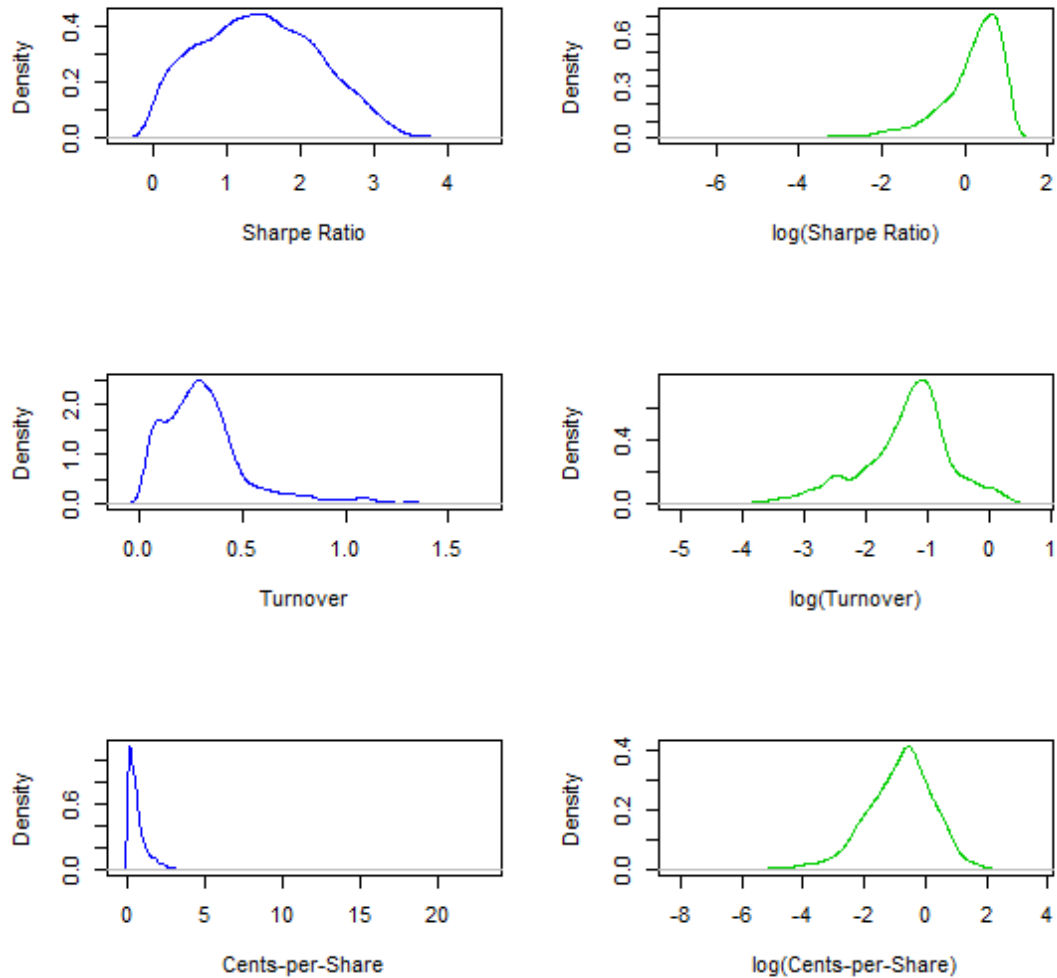

**Figure 1.** Density (using R function `density()`) plots for the Sharpe ratio $S$, turnover $T$ and cents-per-share $C$ and their logarithms for the 3,636 alphas with positive Sharpe ratio (before removing the outliers) discussed in Section 2.

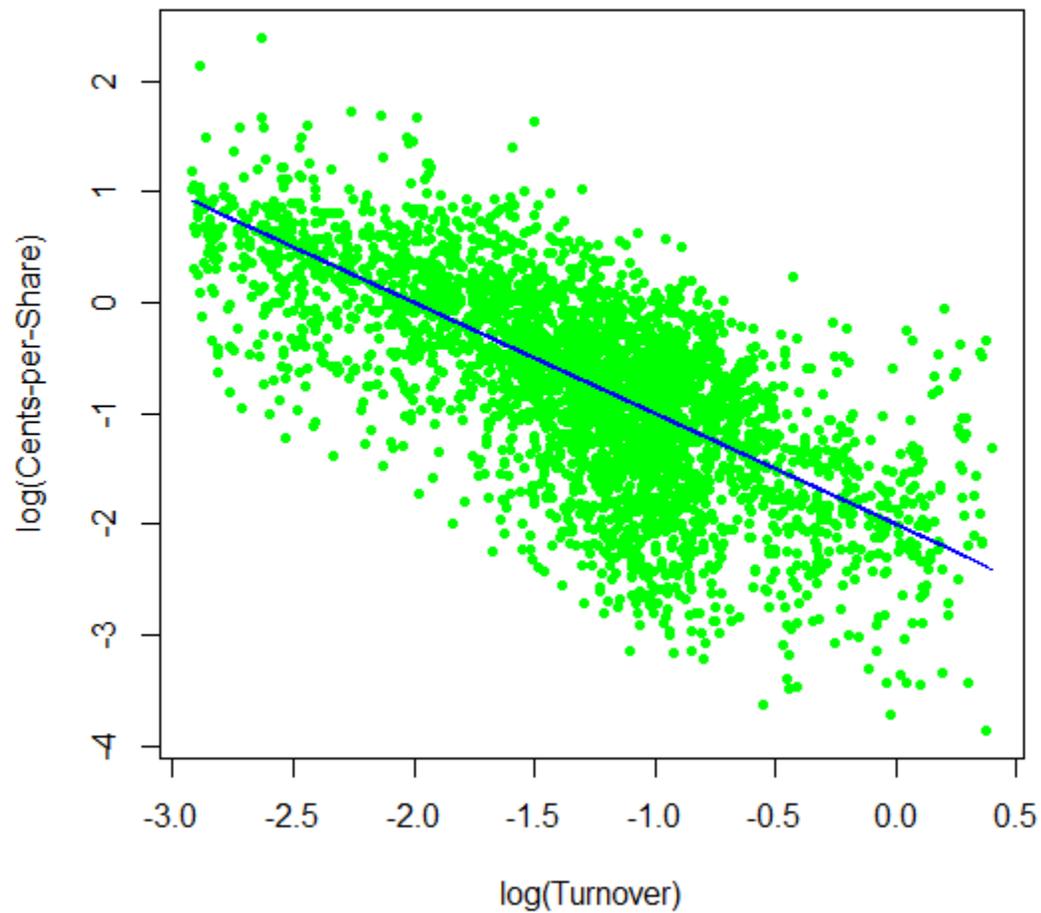

**Figure 2.** Horizontal axis: $\ln(T)$; vertical axis: $\ln(C)$. The dots represent the data points. The straight line plots the function $f(T) = -2 - \ln(T)$. See Subsection 3.1 and Table 2 for details.

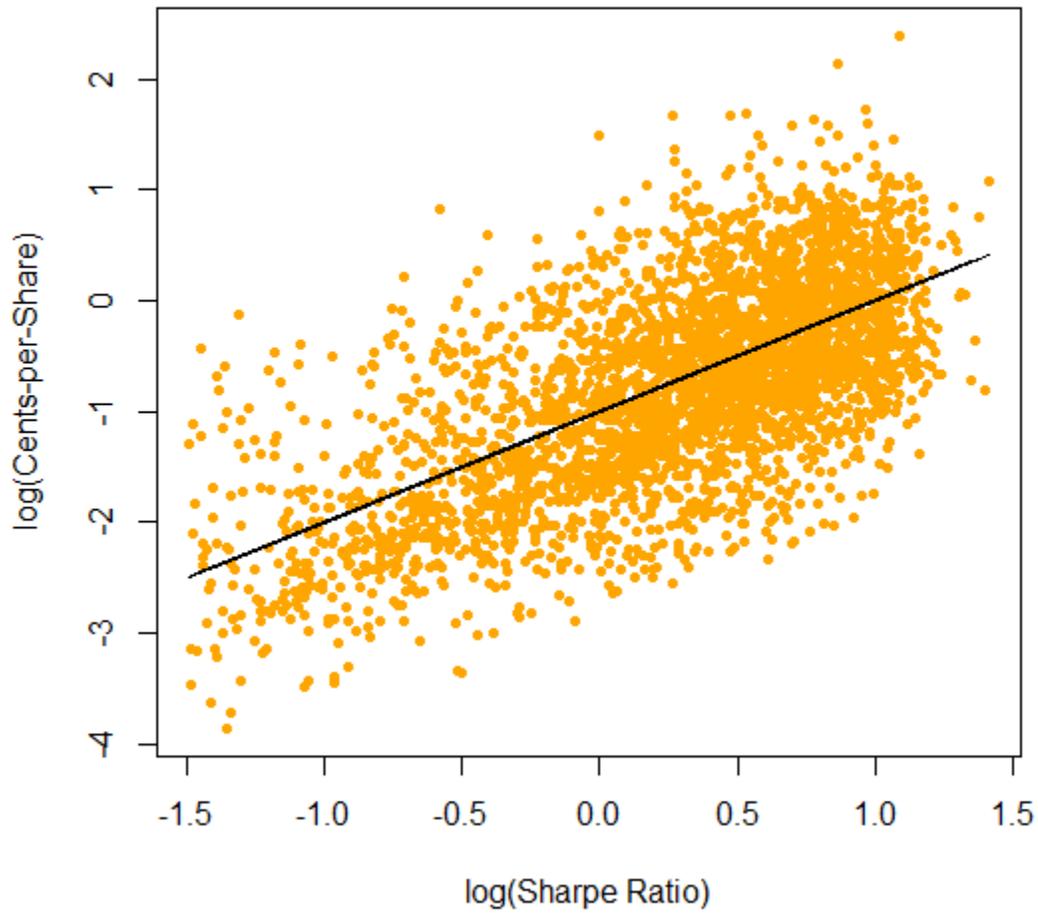

**Figure 3.** Horizontal axis: $\ln(S)$; vertical axis: $\ln(C)$. The dots represent the data points. The straight line plots the function $f(S) = -1 + \ln(S)$. See Subsection 3.2 and Table 3 for details.

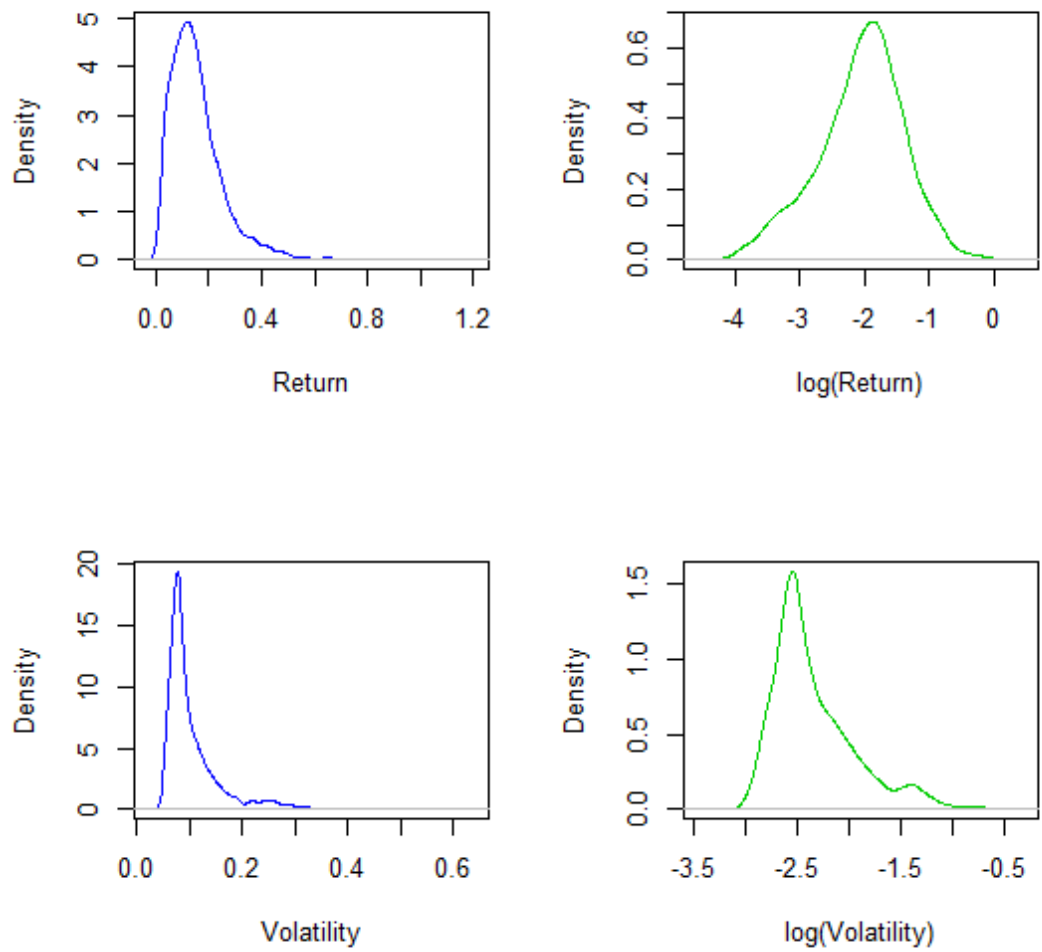

**Figure 4.** Density (using R function `density()`) plots for the return $R'$ and the volatility $\sigma'$ and their logarithms for the 3,289 alphas discussed in Section 2. See Subsection 3.4 for details.

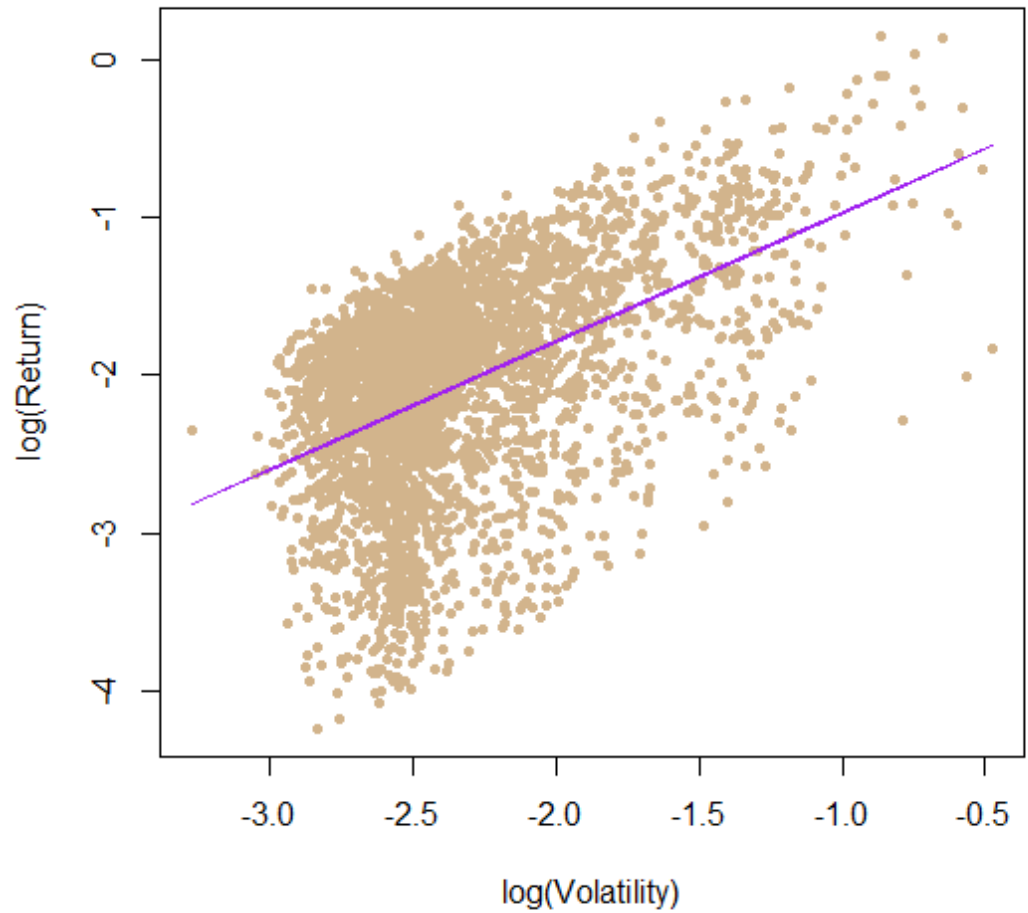

**Figure 5.** Horizontal axis: $\ln(\sigma')$; vertical axis: $\ln(R')$. The dots represent the data points. The straight line plots the function $f(\sigma') = -0.159 + 0.813 \ln(\sigma')$. See Subsection 3.4 and Table 5 for details.

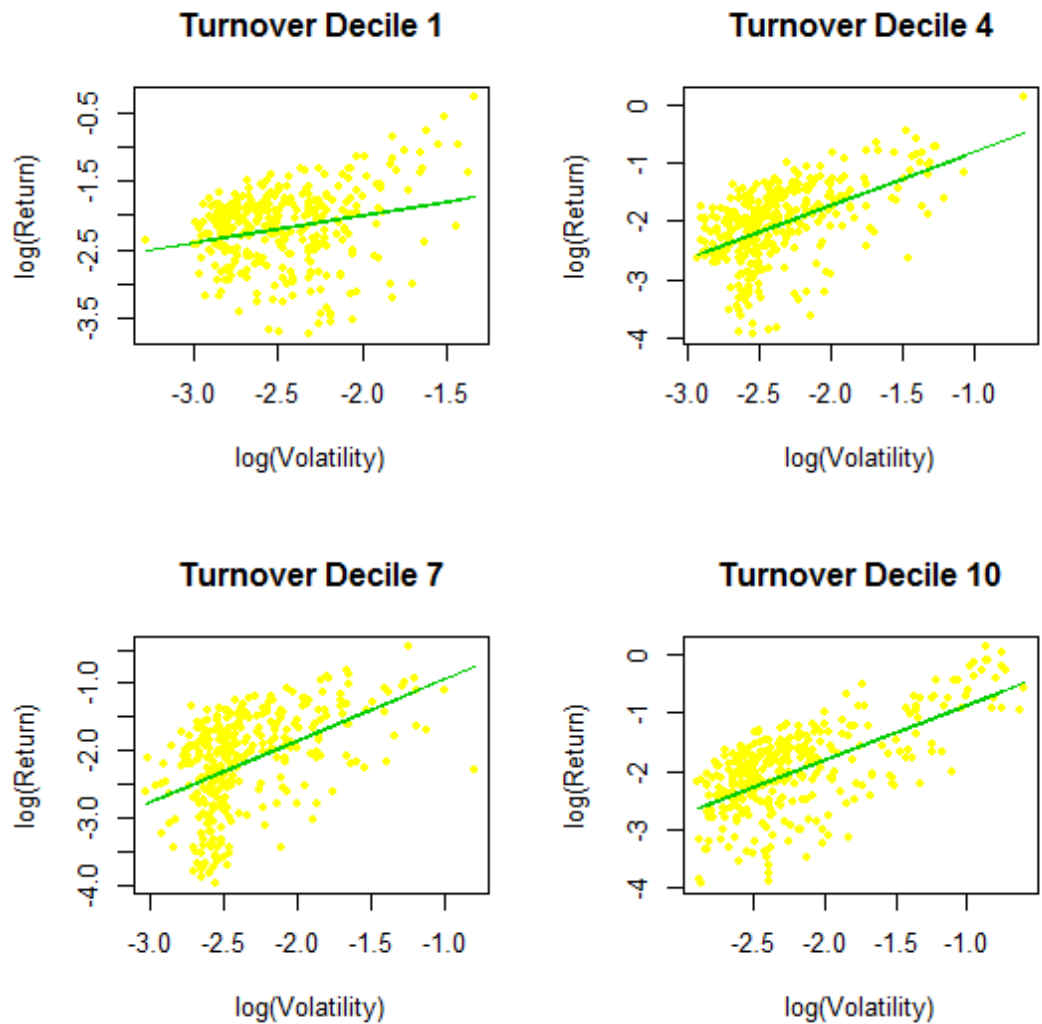

**Figure 6.** Plots for return v. volatility, same as in Figure 5, for the daily turnover $T$ deciles 1, 4, 7 and 10. Horizontal axes: $\ln(\sigma')$; vertical axes: $\ln(R')$. The dots represent the data points. The straight lines plot the corresponding regression fits. See Subsection 3.4 and Table 6 for details.

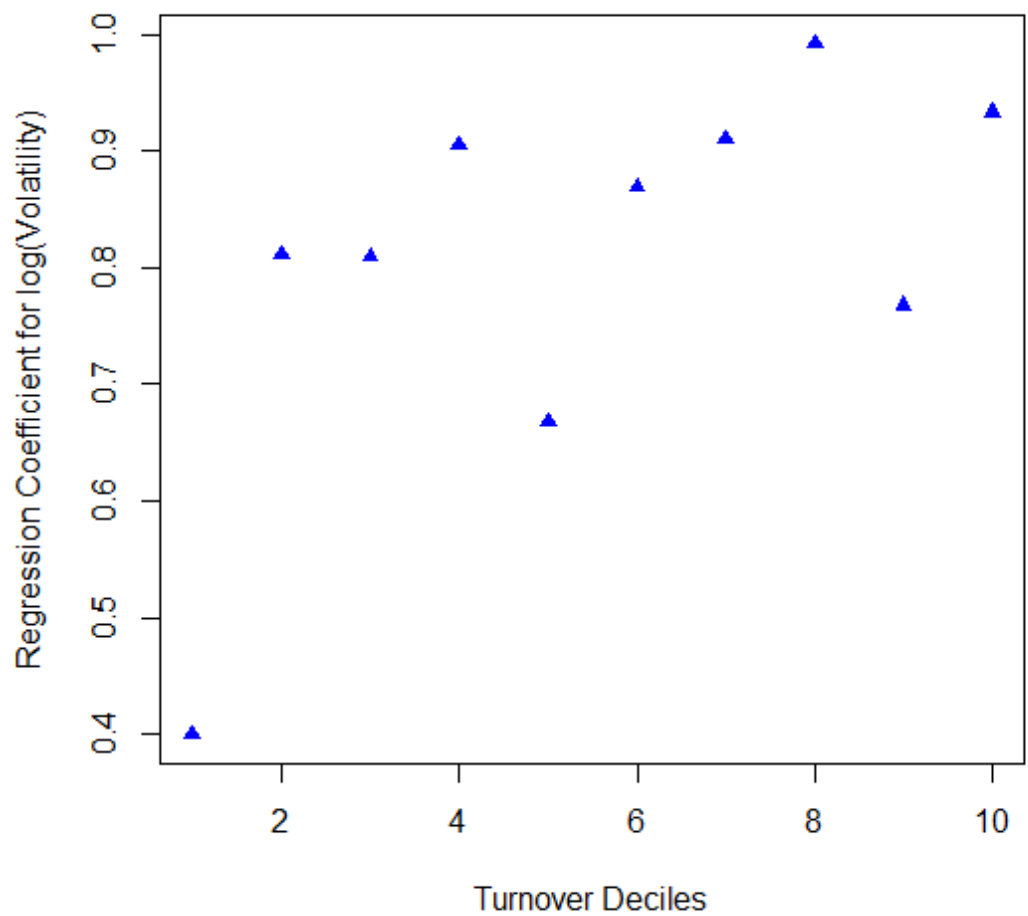

**Figure 7.** Horizontal axis: $T$ deciles; vertical axis: regression coefficient for $\ln(\sigma')$. See Subsection 3.4 and Table 6 for details.